\documentclass[twocolumn,showpacs,preprintnumbers,amssymb,nofootinbib]{revtex4}

\usepackage{graphicx,epsf, epsfig, psfig, amssymb}
\usepackage{bm} 

\def\be{\begin{equation}}
\def\ee{\end{equation}}
\def\beq{\begin{eqnarray}}
\def\eeq{\end{eqnarray}}

\def\ii{{\rm i}}

\def\IL{\relax{\rm I\kern-.18em L}}

\def\f{\frac}

\begin{document}

\title{Quasinormal modes and stability of the rotating acoustic black hole: 
numerical analysis}

\author{Vitor Cardoso}
\email{vcardoso@teor.fis.uc.pt} \affiliation{Centro de F\'{\i}sica
Computacional, Universidade de Coimbra, P-3004-516 Coimbra,
Portugal}

\author{Jos\'e P. S. Lemos}
\email{lemos@fisica.ist.utl.pt} \affiliation{ Centro
Multidisciplinar de Astrof\'{\i}sica - CENTRA, Departamento de
F\'{\i}sica, Instituto Superior T\'ecnico, 
Universidade T\'ecnica de Lisboa, Av. Rovisco Pais 1,
1049-001 Lisboa, Portugal}

\author{Shijun Yoshida}
\email{shijun@waseda.jp} \affiliation{Science and Engineering, 
Waseda University, Okubo, Shinjuku, 
Tokyo 169-8555, Japan}

\date{\today}

\begin{abstract}

The study of the quasinormal modes (QNMs) of the 2+1 dimensional
rotating draining bathtub acoustic black hole, the
closest analogue found so far to the Kerr black hole, 
is performed. Both the real and imaginary parts of the 
quasinormal (QN) frequencies as a function of the rotation 
parameter $B$ are found through 
a full non-linear numerical analysis. Since there is no change 
in sign in the imaginary part of the frequency as $B$ is 
increased we conclude that the 2+1 dimensional
rotating draining bathtub acoustic black hole is stable 
against small perturbations. 

\end{abstract}


\maketitle

\section{Introduction}

A guitar string in vacuum vibrates in normal modes. However, a guitar
string immersed in air vibrates truly in quasinormal modes (QNMs) since it
looses energy to sound waves. The same happens to an acoustic black
hole.  Due to energy losses to the medium in sound wave acoustic
perturbations, its modes of vibration are quasinormal (QN). This
QN behavior is expected, as acoustic black holes are real
analogues of true general relativistic black holes which are known to
vibrate in a QN form due to losses in gravitational and other
types of radiation to infinity.

Acoustic black holes and their analogies with true black holes were
found by Unruh by noticing that the equations of motion for sound
waves in a background fluid flow and the equations of motions for a
scalar field in a black hole background are similar \cite{unruh}. The
background fluid flow acts as an effective black hole metric implying
the existence of a sound horizon, the surface where the velocity of
the flow is equal to the velocity of sound in the medium. A number of
phenomena that occurs with true black holes also occurs with acoustic
black holes. For instance, Hawking radiation is now phonon radiation
\cite{jacobson,unruh2,visser,fischervolovik}. Geodesic and causal
structure can also be studied as was done in \cite{barceloetal2004}
where Penrose-Carter diagrams for several effective acoustic
spacetimes were drawn.  Black holes in other analogue systems, like 
condensed matter ones, can also  be studied 
(see \cite{reznik,garay,barceloetal} to name a few).

In the study of true general relativistic black holes 
QNMs are important for a number of reasons. The most
important is that they provide means to identifying black hole
parameters like the mass and angular momentum, since the real and
imaginary part of the QN frequencies, which give the vibration
frequency and the exponential 
damping frequency respectively, depend only on those
parameters. Through their study one further tests the stability
of the system, as any imaginary frequency with the wrong sign 
would mean an exponentially growing mode, rather than damping. 
QNMs have also been connected to the quantization of the
black hole area, where it seems that the highly damped modes, those
that are almost instantaneous, are associated with transitions between
area levels at large quantum numbers (see \cite{vitorthesis2004} for a
review on QNMs and a list of complete 
citations therein). Since QNMs appear naturally 
in general relativistic black holes, 
they also should appear in their analogues, the acoustic black holes.

The study of QNMs of the 2+1 dimensional rotating draining bathtub
acoustic black hole \cite{visser,basak}, the closest analogue found so
far to the Kerr black hole (see \cite{visserkerr} for developments 
on this subject), was initiated in \cite{berticardosolemos}
through the use of a WKB approximation, valid for small black hole 
rotation. Various interesting points besides QNM behavior, like 
late-time tails and superradiance, were also discussed. In relation 
to large damped QNMs it was found that there are no asymptotic
QN frequencies, a puzzling result if connected to the area quantization 
issue. 
Moreover, the WKB scheme used in \cite{berticardosolemos} 
gave indications that, due to a possible change 
in the sign of the imaginary part of the frequency
$\omega$, this 2+1 acoustic black hole could 
become unstable at large
rotation parameter $B$, or at large azimuthal number $m$.
Recent work on this \cite{lepe}, making use of
a matching procedure, is also not appropriate to
give a convincing stability proof. 
Is the Im$\,\omega$ part lower, equal or greater than zero 
for $B$ or $m$ large? 
In this paper 
we make a full numerical analysis to settle this issue. 
We show that the instability never sets in.

\section{Formalism}

The draining bathtub model is a $2+1$
dimensional flow with a singular vortex-like 
sink (or source) at the origin.
The acoustic metric describing the
propagation of sound waves in this fluid flow is
\cite{visser}:
\begin{eqnarray}
ds^2\!\!&=&\!\!
-\left (c^2-\frac{A^2+B^2}{r^2} \right )dt^2 \nonumber \\
& & +\frac{2A}{r}drdt-2Bd\phi dt+dr^2+
r^2d\phi^2\,.\nonumber \\
& &
\label{metric1}
\end{eqnarray}
Here $A$ and $B$ are arbitrary real positive constants related to the
radial and angular components of the background fluid velocity:
\be
{\vec v}=\frac{-A {\vec r}+B {\vec\phi}}{r}\,, 
\ee
where ${\vec r}$ and ${\vec\phi}$ are orthogonal unit basis vectors 
pointing along the axes. 
It is however better to work with a more transparent metric.
Some physical properties of our
draining bathtub metric are more apparent if we cast the metric in a
Kerr-like form performing the following coordinate transformation
(see \cite{basak,berticardosolemos}):
\begin{eqnarray}
dt&\rightarrow& d\tilde{t}=dt-\frac{Ar}{r^2c^2-A^2}dr\\
d\phi&\rightarrow& d\tilde{\phi}=d\phi-\frac{BA}{r(r^2c^2-A^2)}dr\,,
\label{coordtransf}
\end{eqnarray}
Then the effective metric takes the form
\begin{eqnarray}
ds^2\!\!&=&\!\!
-\left(1-\frac{A^2+B^2}{c^2r^2} \right)c^2 d\tilde{t}^2  \nonumber \\
& &+ \left(1-\frac{A^2}{c^2r^2} \right )^{-1}dr^2
-2B d\tilde{\phi}d\tilde{t}+r^2d\tilde{\phi}^2\,. \nonumber \\
& &
\label{metric2}
\end{eqnarray}
As explained in \cite{berticardosolemos},
this metric and the Kerr metric differ in an important aspect, in that
whereas the rotation for the Kerr black hole is bounded from above,
here it is not, at least in principle. Thus, $B$ could be as large as
desired.

The propagation of a sound wave
in a barotropic inviscid fluid with irrotational flow, 
which is assumed to be the case, is described by
the Klein-Gordon equation $\nabla_{\mu}\nabla^{\mu}\Phi=0$ for a
massless field $\Psi$ in a Lorentzian acoustic geometry.
Separating variables by the
substitution
\be
\Phi(\tilde{t},r,\tilde{\phi})=
 \sqrt{r} \Psi(r)e^{\ii (m\tilde{\phi}-\omega\tilde{t})}\,,
\ee
implies that $\Psi(r)$ obeys the wave equation
\be
\Psi_{,\hat r_* \hat r_*}+Q \Psi=0\,,
\label{waveequation2}
\ee
where the generalized potential is given by 
\be\label{Qdef}
Q\equiv \left(\hat{\omega}-\f{\hat{B}m}{\hat r^2}\right)^2-
\left(\f{\hat r^2-1}{\hat r^2}\right)
\left[\f{1}{\hat r^2}\left(m^2-\f{1}{4}\right)+\f{5}{4\hat r^4}\right]\,.
\ee
The tortoise coordinate is defined by 
\be
\frac{d\hat{r}_*}{d\hat{r}}=\Delta\,,
\ee
where $\Delta\equiv (1-1/\hat{r}^2)^{-1}$. Explicitly,
\be
\hat{r}_*=\hat{r}+\f{1}{2}\log\left|\f{\hat{r}-1}{\hat{r}+1}\right|\,.
\ee
We have also performed the following rescaling: $\hat r=rA/c$,
$\hat{\omega}=\omega A/c^2$, $\hat{B}=B/A$. 
The rescaling effectively sets $A=c=1$ in the original wave equation,
and picks units such that the acoustic horizon $\hat r_H=1$. From now
on we shall omit hats in all quantities. The rescaled wave equation
(\ref{waveequation2}) will be the starting point of our analysis of
QNMs. 
The characteristic QNMs of the rotating acoustic black hole can be
defined in the usual way, imposing appropriate boundary conditions and
solving the corresponding eigenvalue problem. Close to the sound 
horizon we seek solutions of equation (\ref{waveequation2}) behaving as
\begin{equation}
\Psi \sim e^{- \ii\left (\omega-Bm\right)r_* }\,.
\label{bound1}
\end{equation}
Classically, only ingoing waves -- that is, waves falling into the
acoustic black hole -- should be present at the horizon. This means (according
to our conventions on the time dependence of the perturbations) that
we must choose the minus sign in the exponential. At spatial
infinity the solutions of (\ref{waveequation2}) behave as
\begin{equation}
\Psi \sim e^{ \ii\omega r_*}\,.
\label{bound2}
\end{equation}
In this case we require that only outgoing waves (waves leaving the
domain under study) should be present, and correspondingly choose the
plus sign in the exponential. This boundary condition at infinity may
be cause for objections. Indeed, no actual physical apparatus will be
accurately described by these boundary conditions: a real acoustic
black hole experiment will certainly not extend out to
infinity. However, we may imagine using some absorbing device to
simulate the ``purely outgoing'' wave conditions at infinity (for
another example in which an absorbing device modeling spatial infinity
could be required, cf. Section XI of \cite{schutzhold} -- in
particular their Fig. 5). 

The boundary conditions (\ref{bound1}) and (\ref{bound2}) are
satisfied by only a discrete number of frequencies $\omega$, the 
QN frequencies $\omega_{QN}$.
The QN frequencies are in general complex numbers, and the imaginary
part is usually negative, which means that perturbations die
exponentially as time goes on (recall that the time dependence of the
field is $e^{-i\omega t}$, and a negative imaginary part for $\omega$
means exponential decay with time).  If all the QN frequencies have a
negative imaginary part, we say that the system is stable.  Note
also that, for a given $m$ there will in general be an infinite number
of $\omega_{QN}$ satisfying the boundary conditions. We shall order
these $\omega_{QN}$ by imaginary part: the $\omega_{QN}$ having
smallest (in magnitude) imaginary part will be called the fundamental
($n=0$) frequency, the one having the second lowest imaginary part
will be the first overtone $n=1$, and so on.

\section{Numerical procedure}

\subsection{Frobenius expansion}
In order to numerically obtain the QN frequencies, 
in the present investigation, we make use of Leaver's method 
\cite{Le85,Le90}, which is known to yield excellent results.

Let us introduce a new independent radial variable $x$, defined as 
$x=r^{-1}$. In terms of the new  variable $x$, the tortoise 
coordinate is then reduced to 
\begin{equation}
r_*=x^{-1}+\frac{1}{2}\ln(1-x)-\frac{1}{2}\ln(1+x)\,.  
\end{equation}
The perturbation function $\Psi$ may be expanded around the 
horizon as 
\begin{equation}
\Psi=e^{i\omega x^{-1}}\left({1-x\over 1+x}\right)^{-i(\omega-Bm)/2}
\,\sum_{k=0}^\infty 
a_k\left(1-x\right)^k\,, 
\label{expansion}
\end{equation}
where $a_0$ is taken to be $a_0=1$. The expansion coefficients 
$a_k$ in equation (\ref{expansion}) are determined via the 
four-term recurrence relation (it's just a matter of substituting 
expression (\ref{expansion}) in the wave equation 
(\ref{waveequation2})), given by
\begin{eqnarray}
&&\alpha_0a_1+\beta_0a_0=0\,, \nonumber \\
&&\alpha_1a_2+\beta_1a_1+\gamma_1a_0=0\,, \\
&&\alpha_ka_{k+1}+\beta_ka_k+\gamma_ka_{k-1}+\delta_ka_{k-2}=0\,,
\ k=2,3,\cdots, \nonumber
\end{eqnarray}
where 
\begin{eqnarray}
\alpha_k&=& -8 (1+k)\,(1+k+iBm-i\omega) \,, \nonumber \\
\beta_k &=&4 \{1+5k^2+m^2+k\,(5+4iBm-8i\omega) \nonumber \\
& & -4i\omega-4\omega^2+2Bm\,(i+2\omega)\} \,, \nonumber \\
\gamma_k&=&2\,(3-8k^2-4ik\,(Bm-2\omega)) \,, \nonumber \\
\delta_k&=&-3-4k+4k^2 \,. \nonumber
\end{eqnarray}
Since the asymptotic form of the perturbations as 
$r_*\rightarrow\infty$ 
%
$(r_*\rightarrow-\infty)$
%
is written in terms of the variable $x$ as
\begin{equation}
e^{i\omega r_*} \sim e^{i\omega x^{-1}} 
%
\  (e^{-i(\omega-Bm)r_*} \sim (1-x)^{-i(\omega-Bm)/2} )
%
\,,
\label{expwr} 
\end{equation}
the expanded perturbation function $\Psi$ defined by equation 
(\ref{expansion}) automatically satisfy the QNM boundary conditions 
if the power series converges for 
$0\le x\le 1$.
%
Making use of a Gaussian elimination \cite{Le90}, we can reduce the 
four-term recurrence relation to the three-term one, given by
\begin{eqnarray}
&&\alpha'_0a_1+\beta'_0a_0=0\,, \nonumber \\
&&\alpha'_ka_{k+1}+\beta'_ka_k+\gamma'_ka_{k-1}=0\,, 
\ k=1,2,\cdots, 
\end{eqnarray}
where $\alpha'_k$, $\beta'_k$, and $\gamma'_k$ are given in terms of 
$\alpha_k$, $\beta_k$, $\gamma_k$ and $\delta_k$ by 
\begin{eqnarray}
\alpha'_k=\alpha_k,\quad \beta'_k=\beta_k,\quad \gamma'_k=\gamma_k, 
\quad {\rm for\ } k=0,1, 
\end{eqnarray}
and
\begin{eqnarray}
\alpha'_k&=&\alpha_k, \nonumber \\
 \beta'_k&=&\beta_k-\alpha'_{k-1}\delta_k/\gamma'_{k-1}\,, \\
\gamma'_k&=&\gamma_k-\beta'_{k-1}\delta_k/\gamma'_{k-1}\,,\quad 
{\rm for\ }k\ge 2\,. \nonumber  
\end{eqnarray}
Now that we have the three-term recurrence relation for determining the 
expansion coefficients $a_k$, the 
convergence condition for the expansion (\ref{expansion}), namely the 
QNM conditions, can be written in terms of the continued fraction 
as \cite{Le85,Gu67}
\begin{eqnarray}
\beta'_0-\frac{\alpha'_0\gamma'_1}{\beta'_1-}
\frac{\alpha'_1\gamma'_2}{\beta'_2-}\frac{\alpha'_2\gamma'_3}{\beta'_3-}
...\equiv 
\beta'_0-\frac{\alpha'_0\gamma'_1}{\beta'_1-
\frac{\alpha'_1\gamma'_2}{\beta'_2-\frac{\alpha'_2\gamma'_3}{\beta'_3-...}}}
=0 \,,
\label{a-eq}
\end{eqnarray}
where the first equality is a notational definition commonly 
used in the literature for infinite continued 
fractions.
Here we shall adopt such a convention.

%
An analysis of the large $k$ behavior of the expansion coefficients 
$a_k$ shows 
\begin{equation}
\lim_{k\rightarrow\infty}{a_{k+1}\over a_k}=1
\pm{(-2 i\omega)^{1\over 2}\over k^{1\over 2}} 
-{3+4i\omega\over 4k}+\cdots\,. 
\label{aoa}
\end{equation}
The series expansion (\ref{expansion}) will converge uniformly only if 
the sign of the second term in the right hand side of (\ref{aoa}) is 
chosen such that ${\rm Re}(\pm(-2 i\omega)^{1\over 2}) < 0$, which will 
happen only for the QN frequencies $\omega$. In case that the 
convergency of the continued fraction (\ref{a-eq}) is not very good, 
one can use Nollert's technique to avoid this difficulty of the 
convergence \cite{Nollert93}. In Nollert's technique, equation 
(\ref{aoa}) plays an essential role. 
%

\section{Numerical Results}
\begin{figure}
\centerline{\includegraphics[width=7 cm,height=7 cm] {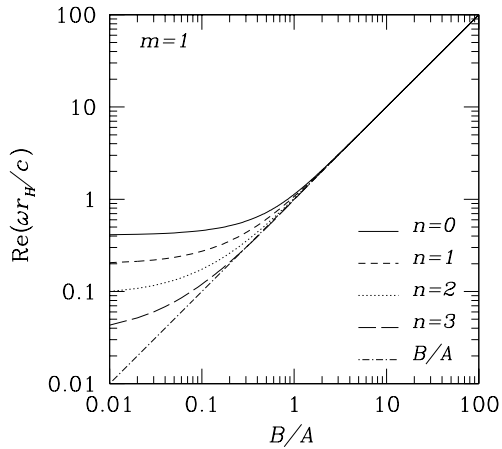}}
\caption{The real part of the QN frequency as a function of the
rotation parameter $B/A$, for several overtones of a $m=1$ mode.
Here, $r_H=A/c$ is the horizon radius. Note how all the several lowest
overtones ``coalesce'' in the high rotation regime, all growing
linearly with $B/A$.}
\label{fig:f1}
\end{figure}
\vskip 1mm

\begin{figure}
\centerline{\includegraphics[width=7 cm,height=7 cm] {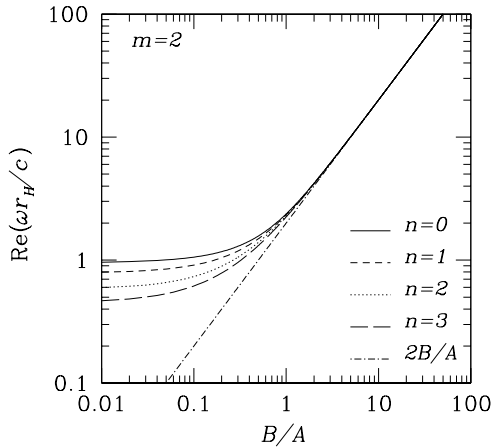}}
\caption{The real part of the QN frequency as a function of
the rotation parameter $B/A$, for several overtones of a $m=2$ mode.
Again, all different overtones have the same behavior for high rotation.} 
\label{fig:f2}
\end{figure}
\vskip 1mm

\begin{figure}
\centerline{\includegraphics[width=7 cm,height=7 cm] {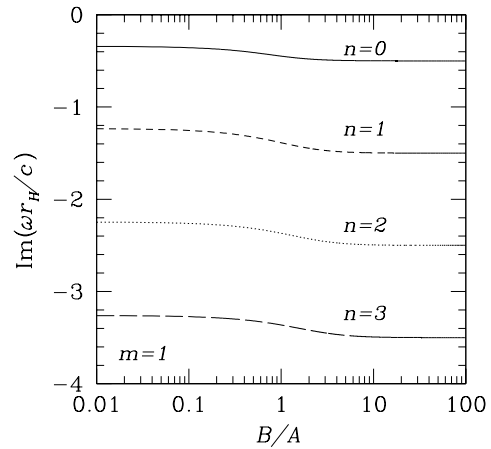}}
\caption{The imaginary part of the QN frequency as a function of
the rotation parameter $B/A$, for several overtones of a $m=1$ mode.
It is clear from this plot that the imaginary part of the QN frequencies
of $m>0$ modes is very insensitive to the rotation of the black hole.
} \label{fig:f3}
\end{figure}
\vskip 1mm

\begin{figure}
\centerline{\includegraphics[width=7 cm,height=7 cm] {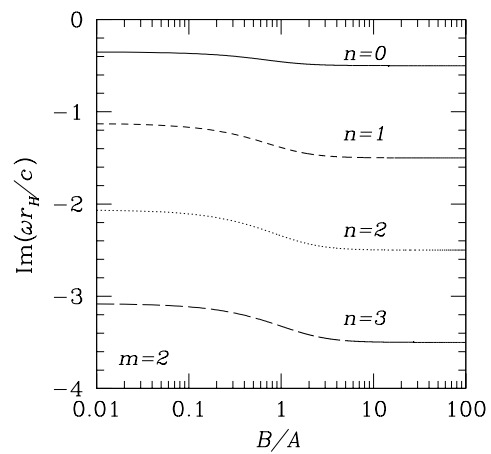}}
\caption{The imaginary part of the QN frequency as a function of
the rotation parameter $B/A$, for several overtones of a $m=2$ mode.} 
\label{fig:f4}
\end{figure}
\vskip 1mm

\begin{figure}
\centerline{\includegraphics[width=7 cm,height=7 cm] {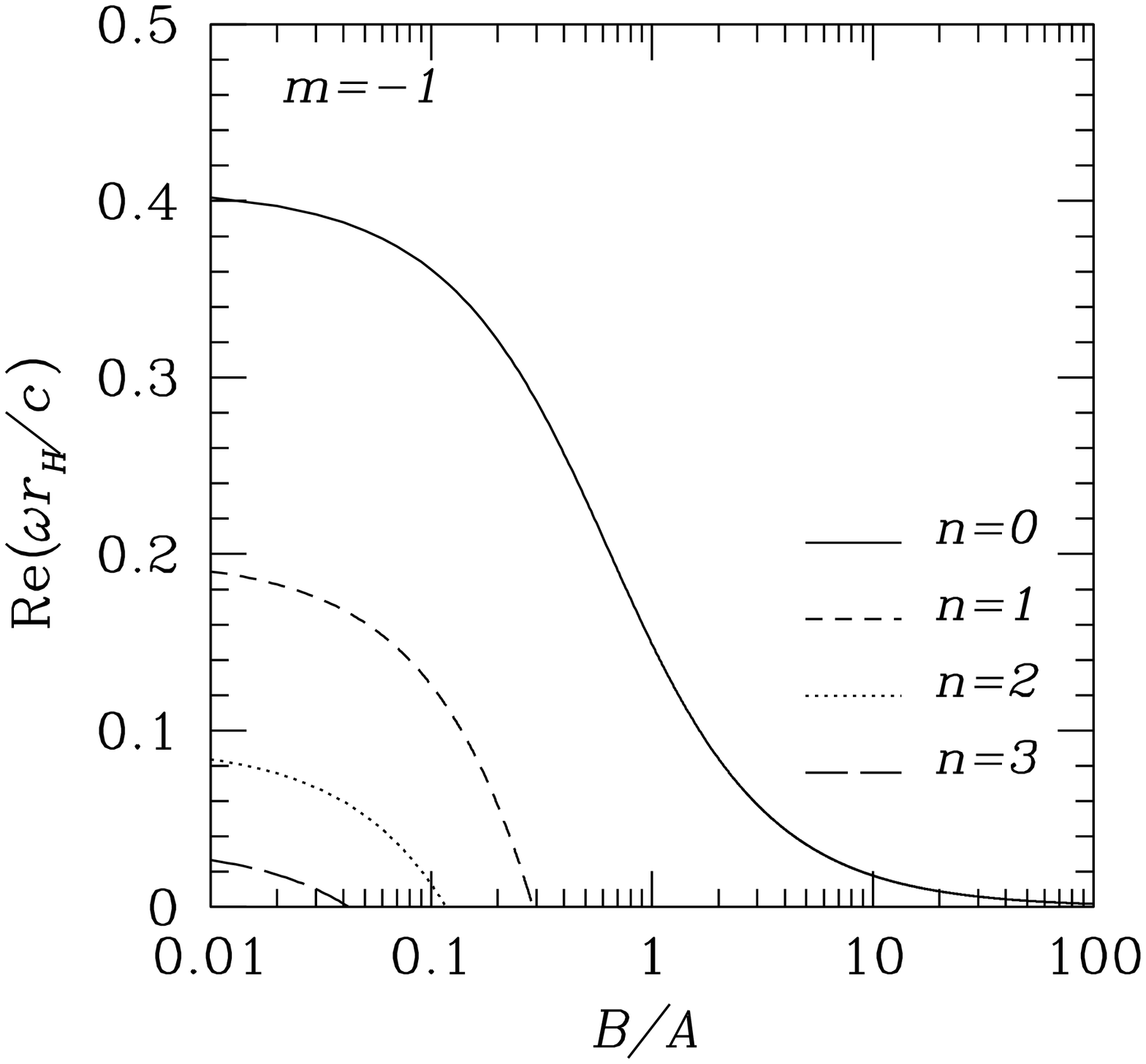}}
\caption{The real part of the QN frequency as a function of
the rotation parameter $B/A$, for several overtones of a $m=-1$ mode.
Notice that for each overtone number $n$ there is a critical rotation
at which the mode crosses the axis, i.e., there is a critical rotation $B/A$
at which the real part of the QN frequency is zero. 
Higher overtones cross the axis at a slower rotation.
We have not been able to follow
the mode beyond this point.} \label{fig:f5}
\end{figure}

\begin{figure}
\centerline{\includegraphics[width=7 cm,height=7 cm] {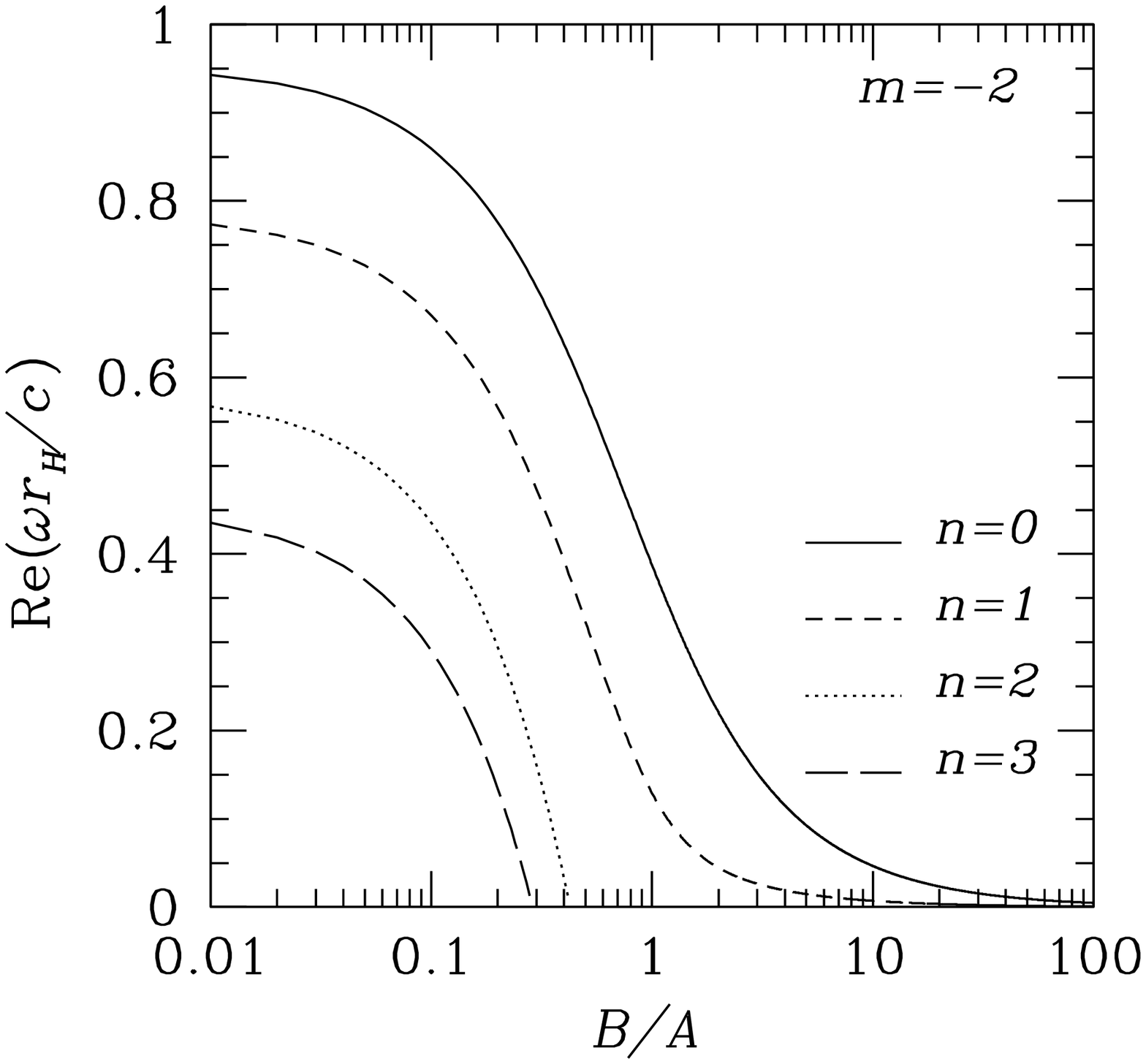}}
\caption{The real part of the QN frequency as a function of
the rotation parameter $B/A$, for several overtones of a $m=-2$ mode.
} \label{fig:f6}
\end{figure}

\begin{figure}
\centerline{\includegraphics[width=7 cm,height=7 cm] {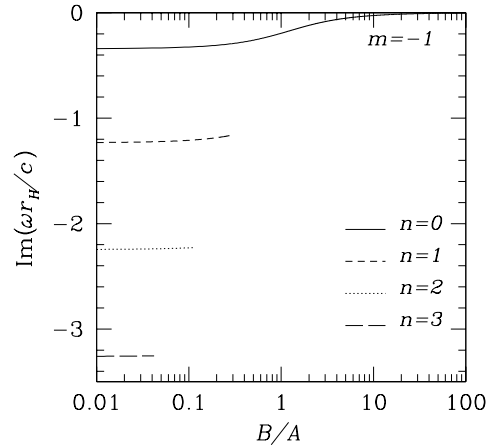}}
\caption{The imaginary part of the QN frequency as a function of the
rotation parameter $B/A$, for several overtones of a $m=-1$ mode.  We
have not been able to follow the modes beyond a certain critical point
(defined as the rotation $B/A$ for which the real part of the QN
frequency is zero- see figs. \ref{fig:f5}-\ref{fig:f6}).
Nevertheless, an judging by the modes we did manage to follow, namely
the fundamental mode, it seems that ${\rm Im}[\omega_{QN}]$ never
crosses the axis, i.e., it is always negative, and therefore the mode
is stable. }
\label{fig:f7}
\end{figure}

\begin{figure}
\centerline{\includegraphics[width=7 cm,height=7 cm] {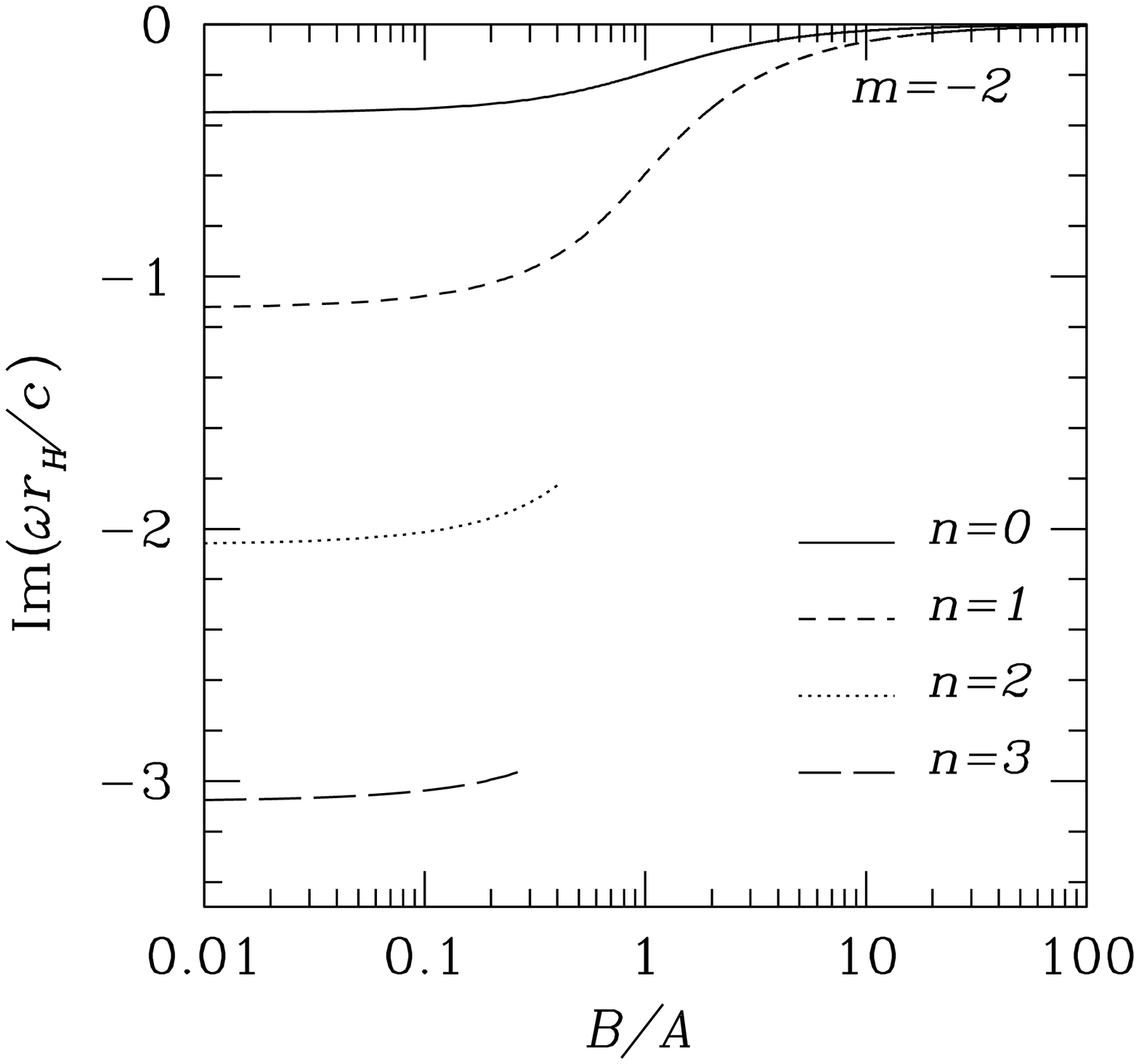}}
\caption{The imaginary part of the QN frequency as a function of
the rotation parameter $B/A$, for several overtones of a $m=-2$ mode.
} \label{fig:f8}
\end{figure}

Our numerical results, which are all consistent with the results
in \cite{berticardosolemos}, are shown in Figs. \ref{fig:f1}-\ref{fig:f8}.

\vskip 1mm

\vskip 1mm

\vskip 1mm

\vskip 1mm

\noindent {\bf (i) $m>0$:}
In Fig. \ref{fig:f1}-\ref{fig:f4} we show results pertaining to
perturbations having positive $m$, i.e., co-rotating waves. 
In Figs. \ref{fig:f1}-\ref{fig:f2} we show the real part of the
QN frequencies for $m=1$ and $m=2$ modes respectively, as a function
of the black hole rotation. Higher $m$
modes follow a similar pattern. 
One can see from these plots that for
low black hole rotation parameter $B$ the different overtones are 
clearly distinguished, but that as the rotation increases they tend to cluster
and behave very similarly. For very large rotation $B$, all the overtones
behave in the same manner, and in this high rotation regime the real part
of the QN frequency scales linearly with the rotation.
Indeed we find that the slope is also proportional to $m$ so that
\be
{\rm Re}[\omega_{QN}] \simeq \frac{mBc^2}{A^2}\quad{\rm as}\ \   
B \rightarrow \infty\,,\quad{\rm for}\ \ m>0
\label{mpos}
\ee
%
We notice that this behavior was already present in the WKB investigation
in \cite{berticardosolemos}.
In Figs. \ref{fig:f3}-\ref{fig:f4} we show the imaginary part of the
QN frequencies as a function of the rotation parameter, for $m=1$
and $m=2$ respectively.
Different overtones have different imaginary parts. 
Note also that for high $B$ the real part of the modes 
coalesce whereas the imaginary part does not.
The magnitude of ${\rm Im}[\omega_{QN}]$ increases
with $B$, which was observed also in the WKB approach \cite{berticardosolemos}.
Thus, as the rotation increases the perturbation dies off quicker.
This also means that the black hole is stable against $m>0$
perturbations, because the imaginary part is always negative.

\noindent {\bf (ii) $m<0$:} 
In Figs. \ref{fig:f5}-\ref{fig:f8} we show results concerning
perturbations having negative $m$, i.e., counter-rotating waves.  The
behavior of the QN frequencies for $m<0$ is drastically different from
the $m>0$ perturbations.  In Figs. \ref{fig:f5}-\ref{fig:f6} we plot
the dependence of ${\rm Re}[\omega_{QN}]$ as a function of the
rotation of the black hole $B$. As $B$ increases the magnitude of the
real part of the QN frequency decreases. 
%
The oscillation frequencies for the fundamental modes, labeled by $n=0$, 
indeed get close to the horizontal axis as $B$ goes to infinity. 
However, we haven't been able to track some overtone 
modes with negative $m$ for very high rotation since, as
can be seen in Figs. \ref{fig:f5}-\ref{fig:f6}, the real part of these
modes eventually change sign.  It is extremely difficult, using the
method employed here, to compute modes having ${\rm Re}[\omega_{QN}]
\sim 0$. Nevertheless, supposing that (as the numerical studies
for the fundamental modes
indicate) the QN frequencies asymptote to zero for very large B, a WKB 
\cite{romanwkb}
analysis shows that $\omega_{QN} \sim k/B$, where $k$ is some
$m$-dependent constant.  The imaginary part of the QN frequencies
behaves in a similar manner, as seen in
Figs. \ref{fig:f7}-\ref{fig:f8}.

\noindent {\bf (iii) $m=0$:} 
For circularly symmetric ($m=0$) modes, our numerical
method shows no sign of convergence.
For $m=0$, the equation (\ref{waveequation2}) can be written in the 
simpler form
\be
\Psi_{,\hat r_* \hat r_*}+(\omega ^2-V) \Psi=0\,,
\label{waveequation3}
\ee
where
\be
V=\left(\f{\hat r^2-1}{\hat r^2}\right)
\left[-\f{1}{4\hat r^2}+\f{5}{4\hat r^4}\right]\,.
\ee
The potential $V$ is not positive definite, and this precludes also
a simple stability proof.

\section{Conclusions}\label{conclusions}
In this paper we have studied numerically the quasinormal modes of the
$(2+1)$-dimensional draining bathtub metric,  which describes a rotating 
acoustic black hole.  Our results indicate that this is a
metric stable against small perturbations, although one would like to
have also an analytical proof of this statement. Notice that this
proof, would most certainly encompass also the the stability of the
usual general relativistic black holes against perturbations of a
charged scalar field;
In fact, for such a field it is possible to show \cite{piran} 
that the generalized potential
goes like $(\omega-eq/r)^2-V$, where $e$ is the scalar field charge, $q$ the
charge of the black hole, and $V$ an $\omega$-independent potential.
This is of the same form as the generalized potential dealt with here,
equations (\ref{waveequation3})-(\ref{Qdef}).
Ever since the first studies on the Kerr geometry (see for example \cite{frolov},
and \cite{cardoso3} for numerical results regarding the QN frequencies of rotating Kerr 
black holes),
one knows that it is extremely difficult to prove stability when $\omega^2-Q$ is not
positive definite (or when it is $\omega$-dependent), and therefore the special
case $Q=(\omega-eq/r)^2-V$ could shed some light on this problem.
To conclude, we would like to draw attention to the fact that the
results presented here are similar in many respects to the results
concerning higher dimensional rotating black holes
\cite{berti}.
\section*{Acknowledgements}
This work was partially funded by Funda\c c\~ao para a
Ci\^encia e Tecnologia (FCT) -- Portugal through project
CERN/FNU/43797/2001.  V.C. acknowledges financial support from FCT
through grant SFRH/BPD/2003. 
S.Y. is supported by the Grant-in-Aid for the 21st Century COE ``Holistic 
Research and Education Center for Physics of Self-organization Systems'' 
from the ministry of  Education, Science, Sports, Technology, and Culture 
of Japan.  


\end{document}